\documentclass[numberedappendix]{emulateapj}

\usepackage{natbib}
\usepackage{graphicx}
\usepackage{amsbsy, amsmath}

\begin{document}

\title{Radiative Rayleigh-Taylor instabilities}

\author{Emmanuel Jacquet}
\affil{Laboratoire de Min\'{e}ralogie et Cosmochimie de Mus\'{e}um (LMCM), CNRS \& Mus\'{e}um National d'Histoire Naturelle, UMR 7202, 57 rue Cuvier, 75005 Paris, France.}
\email{ejacquet@mnhn.fr}  

\author{Mark Krumholz}
\affil{Department of Astronomy, University of California, Santa Cruz, CA 95064}
\email{krumholz@ucolick.org}

\begin{abstract}
We perform analytic linear stability analyses of an interface separating two stratified media threaded by a radiation flux, a configuration relevant in several astrophysical contexts. We develop a general framework for analyzing such systems, and obtain exact stability conditions in several limiting cases.
In the optically thin, isothermal regime, where the discontinuity is chemical in nature (e.g.\ at the boundary of a radiation pressure-driven H~\textsc{ii} region), radiation acts as part of an effective gravitational field, and instability arises if the effective gravity per unit volume toward the interface overcomes that away from it. In the optically thick ``adiabatic'' regime where the total (gas plus radiation) specific entropy of a Lagrangian fluid element is conserved,for example at the edge of radiation pressure-driven bubble around a young massive star, we show that radiation acts like a modified equation of state, and we derive a generalized version of the classical  Rayleigh-Taylor stability condition.
\end{abstract}

\section{Introduction}
The superposition of a dense fluid above a lighter one in a gravitational field is prone to the well-known Rayleigh-Taylor instability \citep[e.g.][]{Chandrasekhar1981}: Any corrugation of the interface between them will grow exponentially, as fingers of the heavier fluid sink in the more buoyant one.
The Rayleigh-Taylor instability and related processes have found applications in various astrophysical settings, such as the expansion of supernova remnants \citep[e.g.][]{Ribeyreetal2004} (where inertial acceleration plays the role of the gravitational field), the interiors of red giants, subject to thermohaline mixing \citep[e.g.][]{CharbonnelLagarde2010}, or interstellar gas clouds pushed above the galactic plane \citep[e.g.][]{Zweibel1991}.

One can envision several Rayleigh-Taylor-like configurations of astrophysical interest where radiation is important for both energetics and dynamics. For instance, during massive star formation, radiation pressure overcomes gravity and causes the formation of bubbles of rarefied matter around the central star(s). Since they are overlain by denser infalling gas, they may be prone to Rayleigh-Taylor instabilities, potentially aiding continued accretion \citep{Krumholzetal2009}. Another astrophysical setting of relevance could be the interface between an H~\textsc{ii} region and its neutral shell. Stellar photons are absorbed in the H~\textsc{ii} region and exert a force toward the interface that acts like an effective gravitational field. In sufficiently dense HII regions driven by sufficiently massive stars this radiation force can be very large \citep{KrumholzMatzner09, Draine2010}, potentially destabilizing the shell of swept-up material. Finally, in the same vein, the radiation force could be significant in shaping Rayleigh-Taylor instabilities in supernova explosions. 

Radiative Rayleigh-Taylor instabilities are not a new subject. \citet{MathewsBlumenthal1977} studied the stability of surfaces and slabs of fully ionized plasmas and found instability for optically thin clouds at their far side and optically thick ones (using the Boussinesq approximation) with significant amount of neutral gas, or pushed at the illuminated side. \citet{Krolik1977} studied the global stability of a constant-density slab under the Boussinesq approximation, and found, in the absence of gravity, instability of short-wavelength perturbations if radiative acceleration correlates positively with total optical depth; inclusion of gravity induced a transition back to the classical Rayleigh-Taylor result. 

A noteworthy related, albeit qualitatively different instability was studied by \citet{BlaesSocrates2003} in the optically thick regime. They performed a local radiative magnetohydrodynamics stability analysis of a stratified equilibrium, and found radiation to overstabilize acoustic disturbances for high enough background flux. Radiation slips into rarefied regions giving rise to buoyant ``photon bubbles''. In the absence of magnetic fields, the instability criterion requires the specific opacity to have an explicit dependence on the density or the temperature.  

In this study, we investigate the role of radiation in the linear stability of a single interface between two media, ignoring magnetic fields and chemical processes as well as the structure of the interface. We present general frameworks in the optically thin and optically thick regimes, before giving analytical solutions in limiting cases. In \S 2, we will review the fundamental equations and outline the model and a few generalities, while \S 3 applies our formalism to the standard (non-radiative) Rayleigh-Taylor instability to illustrate how it works. In \S 4, we focus on the optically thin regime, and in particular the isothermal limit, while \S 5 will be devoted to the optically thick regime, and in particular the adiabatic approximation, whereby the total (gas plus radiation) specific entropy is conserved for a Lagrangian fluid element. In \S 6, we conclude.

\section{Generalities}
\subsection{Equations of Radiation Hydrodynamics}  

We begin by reviewing the fundamental equations of radiation hydrodynamics (RHD). Beforehand, a few words on notation: scalars will be written in italics (e.g.\ $a$), vectors in bold (e.g.\ $\mathbf{F}$) and higher-rank tensors in bold calligraphy (e.g.\ $\mathbf{\mathcal{T}}$).  The product of two tensors $\mathbf{\mathcal{U}}$ and $\mathbf{\mathcal{V}}$ is written $\mathbf{\mathcal{U}}\mathbf{\mathcal{V}}$; their contraction is denoted by a dot for a single index ($\mathbf{\mathcal{U}}\cdot\mathbf{\mathcal{V}}$) and a colon for two indices ($\mathbf{\mathcal{U}}:\mathbf{\mathcal{V}}$). Quantities evaluated in the frame comoving with the fluid will be given a subscript 0.

In the nonrelativistic and inviscid limits, the RHD equations are given by \citep{MihalasWeibelMihalas1984}:
\begin{eqnarray}
\frac{\partial \rho}{\partial t} + \nabla\cdot\left(\rho \mathbf{v}\right) & = & 0
\label{continuity-gas}
\\
\rho\frac{D\mathbf{v}}{Dt} & = & \mathbf{G_0}-\nabla P_g +\rho\mathbf{g}
\label{momentum-gas}
\\
\frac{\partial u_g}{\partial t} +\nabla\cdot\left(u_g\mathbf{v}\right) & = & -P_g\nabla\cdot\mathbf{v} + cG_0^0,
\label{internal-energy-gas}
\end{eqnarray}
for the gas (mass conservation, momentum and internal energy equation), and:
\begin{equation}
\frac{\partial E_r}{\partial t}+\nabla\cdot\mathbf{F}=-cG^0
\label{energy-radiation}
\end{equation}
\begin{equation}
\frac{1}{c^2}\frac{\partial\mathbf{F}}{\partial t}+\nabla\cdot\mathcal{P}_r=-\mathbf{G}
\label{momentum-radiation}
\end{equation}
for the radiation (energy and momentum equation).
Here $\rho$, $P_g=\rho a^2$, $u_g=P_g/(\gamma-1)$ are the gas density, pressure and internal energy per unit volume, respectively, with $a=\sqrt{k_BT/m}$ the isothermal sound speed, and $\mathbf{g}=-\nabla\phi$ is the gravitational acceleration with $\phi$ the potential. $E_r$, $\mathbf{F}$, $\mathcal{P}_r$ are the energy density, energy flux vector and pressure tensor of the radiation field.

The rate of 4-momentum transfer from radiation to matter per unit space-time volume $dVdt$ (or, minus the 4-divergence of the radiation energy-momentum tensor) $\mathcal{G}$, 
evaluated in the comoving frame, assuming that the gas is in Local Thermodynamic Equilibrium (LTE), and that scattering is isotropic, is given by:
\begin{equation}
\mathcal{G}_0= 
\left(
\begin{array}{c}
G^0_0 \\
\mathbf{G}_0
\end{array}
\right) =
\left(\begin{array}{c}
\rho(\kappa_JE_{r0}-\kappa_PaT^4) \\ \frac{\kappa_F\rho}{c}\mathbf{F}_0
\end{array}\right),
\end{equation}
where $\kappa_J$ and $\kappa_P$ are frequency-integrated absorption opacity means weighted against the spectral energy distribution (SED) of the radiation and a Planckian at the gas temperature $T$, respectively, and $\kappa_F$ is the flux mean (with both absorption and scattering contributions).

When coupling of the radiation to the gas (through the latter term) is significant, it is useful to rewrite the radiation equations in terms of the comoving frame energy density $E_{r0}=E_r-2\mathbf{v}\cdot\mathbf{F}/c^2$, radiative flux $\mathbf{F_0}=\mathbf{F}-(E_r+\mathcal{P}_r)\mathbf{v}$ and radiation pressure tensor $\mathcal{P}_{r0}=\mathcal{P}_r-(\mathbf{F}\mathbf{v}+\mathbf{v}\mathbf{F})/c^2$ (equations 95.87 and 95.88 of \citet{MihalasWeibelMihalas1984}):
\begin{equation}
\frac{\partial E_{r0}}{\partial t}+\nabla\cdot\left(E_{r0}\mathbf{v}+ \mathbf{F_0}\right)+\mathcal{P}_{r0}:\nabla\mathbf{v}+2\frac{\mathbf{a}\cdot\mathbf{F_0}}{c^2}=-cG_0^0
\label{energy-radiation-comoving}
\end{equation}
\begin{eqnarray}
-\mathbf{G_0} & = & \frac{1}{c^2}\frac{D}{D t}\mathbf{F}_0+\nabla\cdot \mathcal{P}_{r0}+\left(\mathbf{F_0}\cdot\nabla\right)\frac{\mathbf{v}}{c^2}+\frac{\nabla\cdot\mathbf{v}}{c^2}\mathbf{F}_0\nonumber\\
&&{}+\left(E_{r0}+\mathcal{P}_{r0}\right)\frac{\mathbf{a}}{c^2},
\label{momentum-radiation-comoving}
\end{eqnarray}
where it is (generally) safe, for $v\ll c$, to drop all the terms containing $\mathbf{a}\equiv D\mathbf{v}/Dt$ as well as $\left(\mathbf{F_0}\cdot\nabla\right)\mathbf{v}/c^2$ and $(\nabla\cdot\mathbf{v}/c^2)\mathbf{F_0}$. 
Note that this system of equations needs a closure, which will be obtained through various approximations depending on the regime considered in the next sections.

If we sum equations (\ref{momentum-gas}) and (\ref{momentum-radiation}), we obtain the total momentum equation (equation 94.10b of \citet{MihalasWeibelMihalas1984}): 
\begin{equation}
\frac{\partial}{\partial t}\left(\rho\mathbf{v}+\frac{1}{c^2}\mathbf{F}\right)+\nabla\cdot\left(\rho\mathbf{v}\mathbf{v}+\mathcal{P}_r+P_g\mathcal{I}_3\right)=-\rho\nabla\phi,
\label{total-momentum}
\end{equation}
where $\mathcal{I}_3$ is the $3 \times 3$ identity matrix, and we have followed
 \citet{MihalasWeibelMihalas1984} in dropping the term $-G_0^0\mathbf{v}/c$ on the right-hand-side as non-dominant in flows with $v\ll c$.

Equations (\ref{internal-energy-gas}) and (\ref{energy-radiation-comoving}) can be summed to yield (equation 16 of \citet{Buchler1979}):
\begin{equation}
\frac{D E_{\mathrm{tot}}}{Dt}+\nabla\cdot\mathbf{F_0}+\mathcal{H}_{\mathrm{tot}}:\nabla\mathbf{v}=0,
\label{total-internal-energy}
\end{equation}
with $E_{\mathrm{tot}}\equiv E_{r0}+u$ and $\mathcal{H}_{\mathrm{tot}}\equiv E_{\mathrm{tot}}\mathcal{I}_3+\mathcal{P}_{\mathrm{tot}}$, where $\mathcal{P}_{\mathrm{tot}}=P_g\mathcal{I}_3+\mathcal{P}_r$. Yet another useful form of the total energy equation can be obtained by adding the scalar product of equation (\ref{momentum-gas}) with $\mathbf{v}$ (equation 18 of \citet{Buchler1979}):
\begin{eqnarray}
0&=&
\frac{\partial}{\partial t}\left[\rho\left(\frac{\mathbf{v}^2}{2}+\phi\right)+E_{\mathrm{tot}}\right]
+\nabla\cdot{}
\nonumber\\
&&\;\left\{\left[\rho\left(\frac{\mathbf{v}^2}{2}+\phi\right)+E_{\mathrm{tot}}\right]\mathbf{v}
+\mathbf{F_0}+\mathcal{P}_{\mathrm{tot}}\cdot\mathbf{v}\right\},
\label{total-energy}
\end{eqnarray}
where we have assumed the gravitational potential to be static.

\subsection{Model and Linear Stability Formalism}

We consider a plane-parallel background configuration, 
consisting of two semi-infinite media separated by an interface at $z=0$, with medium 1 overlying medium 2 (which one might generally think of as being more rarefied). Throughout this study, we will ignore the width of the discontinuity, and we allow no flow accross it. The system is subject to a constant and uniform external gravitational field (or, equivalently, an inertial acceleration) $\mathbf{g}=-g\mathbf{e}_z$ and is threaded by a radiative flux $\mathbf{F}$, which in equilibrium is independent of $z$ and vertical. For the astrophysical applications considered here, both gravitation and radiation fields may be thought of as being caused by a radiation source such as a massive star located at $z=-\infty$. 

The system of dynamical equations written in the preceding subsection, when supplemented by equations of state and appropriate closures, may be cast in the form
\begin{equation}
i\frac{\partial\psi}{\partial t}=H(\psi),
\end{equation}
where $\psi$ is a vector of the different fields (here, physical quantities as functions of spatial location) evolved in time by the (nonlinear) operator $H$. The equilibrium configuration $\psi_{\rm eq}$ then satisfies $H(\psi_{\rm eq})=0$. Considering a perturbation $\delta\psi\equiv\psi-\psi_{eq}$, we have, to linear order
\begin{equation}
i\frac{\partial\delta\psi}{\partial t}=dH_{\psi_{\rm eq}}(\delta\psi),
\end{equation}
with $dH_{\psi_{\rm eq}}$ the (linear) differential of $H$ at $\psi_{\rm eq}$. 
 The problem now amounts to finding the eigenmodes of $dH_{\psi_{\rm eq}}$, since if $dH_{\psi_{\rm eq}}(\delta\psi(0))=\omega\,\delta\psi(0)$, $\delta\psi(t)=e^{-i\omega t}\delta\psi(0)$. If $\mathrm{Im}(\omega)>0$, the perturbation grows and linear instability is declared.
 We therefore are interested in Eulerian perturbations whose space-time dependence, for any quantity $Q(x,z,t)$, is given by
\begin{equation}
\delta Q(x,z,t)=\delta\hat{Q}(z)e^{i(kx-\omega t)},
\end{equation}
where the Fourier dependence in $x$ (whereby we orient the axes to have $k$ positive) is motivated by the plane parallel nature of the background equilibrium. Since no perturbed vector quantity has a component perpendicular to both $\mathbf{e}_x$ and $\mathbf{e}_z$, the \textit{linear} problem is 2D. 

The eigenvalue problem now reduces to a set of coupled ODEs, supplemented by a set of relationships with no derivatives in $z$, and the former may be cast in the form:
\begin{equation}
\frac{d\delta\hat{\psi}}{dz}=A(z) \cdot \delta\hat{\psi},
\end{equation}
where the linear operator $A$, in our problem, depends on $z$ only through the background quantities, in turn completely determined by their values at $z=0^\pm$ and the values of $g$ and $F_z$ (the $z$ component of the radiation flux) from the equilibrium equations.

Since each solution to this set of ODEs corresponds to a set of perturbations in the space $z=0^\pm$, it should \textit{in principle} be possible to analyse stability conditions as a function solely of quantities evaluated at the interface (rather than integrals, as in \citet{Krolik1977}, but he was considering an upper boundary for the cloud). 

\subsection{Boundary conditions}

Up to this point, nothing distinguishes our problem mathematically from a stability analysis of an infinite, single medium, be the analysis global or local in nature. 
The distinguishing characteristic of the interface problem is the boundary conditions which select the relevant ($z$ dependence of the) eigenfunctions, on which we now focus. 

First, we consider media that are unbounded on either side of the interface, so we require our modes not to blow up as $z$ goes to $\pm\infty$. Thus we are focusing on ``local'' instabilities at the interface, rather than global ones on a cloud scale as in \citet{Krolik1977}. This requires
\begin{equation}
\lim_{z \to\pm\infty}\delta\hat{\psi}(z)=0.
\end{equation}

We next investigate the continuity conditions at the interface. Let ${\boldsymbol \xi}(x,z,t)$ be the Lagrangian displacement of the fluid element that is at position $(x,z)$ in the unperturbed state. We also denote, for any quantity $Q(x,z,t)$ the Lagrangian perturbation by
\begin{equation}
\Delta Q=\delta Q+{\boldsymbol \xi}\cdot\nabla Q,
\end{equation}
 The usual kinematic relationship
\begin{equation}
\Delta\mathbf{v}=\frac{D{\boldsymbol \xi}}{Dt}
\end{equation}
reduces, for the Fourier dependence adopted for our solutions and the zero-velocity background, to $\delta\mathbf{v}=-i\omega{\boldsymbol \xi}$.
Since there is no flow accross the interface, $\xi_z(x,0,t)$ represents the vertical displacement of the boundary between the two fluids. Thus, $\xi_z$ is continuous at the interface.

Now consider a general flux-conservative form equation describing the evolution of the system:
\begin{equation}
\label{general-flux-conservative}
\frac{\partial m}{\partial t}+\nabla\cdot\mathbf{f} = s,
\end{equation}
where $m$ is the conserved quantity, $\mathbf{f}$ is the corresponding flux, and $s$ is a source term. The equations of mass conservation (\ref{continuity-gas}), total momentum conservation (\ref{total-momentum}), and total energy conservation (\ref{total-energy}) are all manifestly of this form. We place ourselves in an inertial frame comoving (at time $t$) with the interface. (Note that for the general considerations we are about to make, it is immaterial whether there is a net flow across it or not). Our purpose here is to find under which conditions the component of the flux $f_{0,z}$ (the 0 subscript referring to the frame chosen) normal to the interface can be considered continuous accross at the interface at $z=\xi_z$.

  Integration of equation (\ref{general-flux-conservative}) accross the interface thickness yields
\begin{eqnarray}
\lefteqn{f_{0,z}(x,\xi_z+\frac{\epsilon}{2},t)-f_{0,z}(x,\xi_z-\frac{\epsilon}{2},t)}
\qquad\qquad\qquad
\nonumber \\
&= & \epsilon\bigg(\langle s\rangle - \frac{\partial\langle m_0\rangle}{\partial t_0}- \frac{\partial\langle f_{0,x}\rangle}{\partial x}\bigg),
\label{integration flux-conservative}
\end{eqnarray}
where $\epsilon$ is the thickness of the interface, and the brackets $\langle ... \rangle$ denote averages accross the interface, i.e., for any function $Q(x,z,t)$
\begin{equation}
\langle Q \rangle \equiv\frac{1}{\epsilon}\int_{\xi_z(x,0,t)-\epsilon/2}^{\xi_z(x,0,t)+\epsilon/2}Q(x,z,t)\, dz,
\end{equation}
which is a function of $x$ and $t$. It is a consequence of the choice of frame and the orientation of $z$ axis normal to the interface\footnote{Actually, the normal to the perturbed interface generally differs from the $z$ axis (defined at equilibrium), such that the relevant component of the flux we should consider is $\left[f_{0,z}-f_{0,x}(\partial\xi_z/\partial x)\right]/\sqrt{1+(\partial\xi_z/\partial x)^2}$, but this does not differ from $f_{0,z}$ to linear order.} that equation (\ref{integration flux-conservative}) has no extra ``boundary term''.

We expect $m_0$, $\mathbf{f}_0$ and $s$ to remain bounded within the interface (although their $z$-\textit{derivatives} may be large) such that their $z$-integrated averages are comparable to their asymptotic values on either side of the interface.\footnote{As regards $s$, it is nonzero only for the momentum equation, where it is proportional to $\rho$ and a fixed gravity; were we including self-gravity, we would even be able to write it as the divergence of a flux  $-\left(\mathbf{g}^2/2-\mathbf{g}\mathbf{g})\right)/4\pi G$ so that no source term would be present.} Therefore, it is already qualitatively clear that the continuity of $f_{0,z}$ will be verified if $\epsilon$ is ``small enough''.

In order to be more quantitative, we note that
\begin{eqnarray}
\lefteqn{f_{0,z}(x,\xi_z+\frac{\epsilon}{2},t)-f_{0,z}(x,\xi_z-\frac{\epsilon}{2},t)}
\nonumber \\
& = &
\left[\Delta f_{0,z}\right]^1_2 + f_{0,z,\mathrm{eq}}(x,\frac{\epsilon_{\mathrm{eq}}}{2},t)-f_{0,z,\mathrm{eq}}(x,-\frac{\epsilon_{\mathrm{eq}}}{2},t)
\nonumber\\
& = & \left[\delta f_{0,z}+s_{\mathrm{eq}}\xi_z \right]^1_2+\langle s_{\mathrm{eq}}\rangle\epsilon_{\mathrm{eq}}
\end{eqnarray}
where subscripts ``eq" refer to the equilibrium, unperturbed value of a quantity, and for any quantity $Q(x,z,t)$ we define
\begin{equation}
\left[Q\right]^1_2\equiv Q(x,0^+,t)-Q(x,0^-,t),
\end{equation}
with $Q(x,0^+,t)$ the value taken by $Q$ in medium 1 at $z=0$ (the value being \textit{extrapolated} if the perturbed interface is actually above $z=0$) and $Q(x,0^-,t)$ that same quantity for medium 2 (extrapolated if the perturbed interface is below $z=0$). We have also used $\partial f_{0,z,\mathrm{eq}}/\partial z = s_{\mathrm{eq}}$.

Since the $\langle s_{\mathrm{eq}}\rangle\epsilon_{\mathrm{eq}}$ term essentially cancels $\langle s\rangle\epsilon$ in the right-hand-side of equation (\ref{integration flux-conservative}), we see that the question of the vertical flux continuity amounts to that of $\Delta f_{0,z}$ (which is actually what we will be using in the stability analyses). Equation (\ref{integration flux-conservative}) may be rewritten as
\begin{equation}
\left[\Delta f_{0,z}\right]^1_2=-\epsilon\left(\frac{\partial \langle m_0 \rangle}{\partial t_0}+\frac{\partial \langle f_{0,x} \rangle}{\partial x}\right)
\end{equation}
So the general condition that our perturbation must satisfy in order to have continuity of $\Delta f_{0,z}$ across the interface is that $k\epsilon\,\delta f_{0,x}\ll \Delta f_{0,z}$ and $\epsilon\omega\,\delta m \ll \Delta f_{0,z}$. If for example, $\delta f_{0,x}\sim\delta f_{0,z}$, we obtain $\epsilon\ll 1/k$, as might have been expected intuitively.

Since application of this boundary condition to the mass conservation equation (\ref{continuity-gas}) does not bring any new information as there is no flow accross the interface, and since we will be making approximations to the energy equations, the sole important application (in this paper) of the above considerations is the $z$ component of the momentum equation (\ref{total-momentum}), where $m=\rho v_z + F_z/c^2$, $\mathbf{f} =\rho v_z\mathbf{v}+ \mathcal{P}_r\cdot \mathbf{e}_z+P_g\mathbf{e}_z$, and $s=-\rho g$. Thus we have $f_{0,z}=\mathcal{P}_{r0}^{zz}+P_g$, and the result will thus read
\begin{equation}
\left[\Delta f_{0,z}\right]^1_2=\left[\delta P_g+\delta \mathcal{P}_{r0}^{zz}-\rho g\xi_z\right]^1_2=0.
\end{equation} 
(From now on, we shall drop the ``eq'' subscripts from the background quantities.) To linear order, we will always have $f_{0,x}=0$, so the only important condition is $\epsilon\omega\, \delta m\ll\rho g \xi_z$. If we take $\delta m \sim \rho\omega\xi_z$, one obtains the condition $\epsilon\ll g/\omega^2$. For $\omega$ of order the classical Rayleigh-Taylor result (rederived in the next section), this amounts to the constraint $k\epsilon\ll 1$, i.e.\ that continuity holds as long as we restrict ourselves to considering perturbations with wavelengths much larger than the thickness of the interface. In the case where radiation forces are important, this is likely to be of order the photon mean free path or the radiation diffusion length, depending on the particular problem we are considering.

\section{The classical Rayleigh-Taylor instability}

We illustrate the above formalism with the classical Rayleigh-Taylor instability, which also provides a benchmark with which the upcoming results can be compared. In this section, we therefore ignore radiation and consider the two media to consist of constant-density (incompressible) fluids (as appropriate for liquids). The perturbed mass conservation (here incompressibility) and Euler equations then read
\begin{eqnarray}
\frac{\partial}{\partial z}\delta v_z+ik\,\delta v_x & =& 0
\label{incompressibility-perturbed-classical}
\\
-i\omega\rho\,\delta v_z + \frac{\partial}{\partial z}\delta P_g & = & 0
\label{Eulerz-perturbed-classical}
\\
-i\omega\rho\,\delta v_x + ik\,\delta P_g & = & 0.
\label{Eulerx-perturbed-classical}
\end{eqnarray}
Solving equation (\ref{Eulerx-perturbed-classical}) for $\delta v_x$ and recalling that $\delta v_z = -i \omega \xi_z$, equations (\ref{incompressibility-perturbed-classical}) and (\ref{Eulerz-perturbed-classical}) yield
\begin{equation}
\frac{d}{dz}
\left[\begin{array}{c}
\hat{\xi}_z\\\delta \hat{P}_g
\end{array}\right]
=
A
\left[\begin{array}{c}
\hat{\xi}_z\\\delta \hat{P}_g
\end{array}\right],
\end{equation}
with:
\begin{equation}
A=
\left[\begin{array}{cc}
0 & \frac{1}{\rho}\left(\frac{k}{\omega}\right)^2\\
\rho\omega^2&0 \\
\end{array}\right]
.
\end{equation}

The matrix $A$ here is independent of $z$ in each medium. In general, for a constant $2\times 2$ matrix $A$ (a circumstance we shall encounter again), the solution for each individual medium (keep in mind $\delta\hat{\psi}$ is \textit{not} continuous accross the interface) may be written as
\begin{equation}
\delta\hat{\psi}(z)= 
\left[
\begin{array}{c}
\hat{\xi}_z \\
\delta \hat{P}_g
\end{array}
\right] =
C_ae^{r_az}\delta\hat{\psi}_a+C_be^{r_bz}\delta\hat{\psi}_b
\label{general-solution}
\end{equation}
where $C_{a,b}$ are two constants of integration and $\delta\hat{\psi}_{a,b}$ are two linearly independent eigenvectors of the matrix $A$, with eigenvalues $r_{a,b}$. In this simple case, the eigenvalues in question are $r_a = k$ and $r_b = -k$. In order for $\hat{\xi}$ and $\delta \hat{P}_g$ not to blow up away from the interface, it is therefore necessary that $C_a = 0$ in the region $z>0$ and $C_b = 0$ in the region $z<0$. $\delta\hat{\psi}$ must thus be an eigenvector of $A$, with eigenvalue $-k$ in medium 1 ($z>0$) and $k$ in medium 2 ($z<0$), respectively.

To obtain the dispersion relation, we apply the boundary conditions at the interface, which in the absence of radiation reads
\begin{equation}
\left[\Delta P_g\right]^1_2=\left[\delta P_g - \rho g \xi_z\right]^1_2=0..
\label{gas-pressure-continuity}
\end{equation}
If we solve for $\delta \hat{P}_g$ as a function of $\hat{\xi}_z$ in the eigenvalue equation for each medium and plug into equation (\ref{gas-pressure-continuity}), we obtain
\begin{equation}
\omega^2=gk\frac{\rho_2-\rho_1}{\rho_2+\rho_1}
\end{equation}
 The instability criterion is thus $\rho_1>\rho_2$ as is well-known. The growth rate of the instability in the limit $\rho_1 \gg \rho_2$ is $\mathrm{Im}(\omega) = \sqrt{gk}$.

\section{The optically thin isothermal regime}

\subsection{Formulation of the equations}

We now consider radiation, first in the optically thin isothermal regime. By optically thin we mean that we can neglect attenuation and treat the radiation flux as constant and unperturbed in each of the two fluids, and by isothermal we mean that each of the fluids is kept at a fixed temperature via its interaction with the radiation. A discontinuity exists only because there is a chemical change at the interface between the two fluids, and possibly a frequency shift in the radiation spectrum at the interface as well (though the total frequency-integrated flux is constant). As a result, the fluid on one side of the interface interacts with radiation differently than fluid on the other side.

One possible astrophysical realization of this situation is an ionization front, where fluid on one side of the interface is ionized and hot, and the radiation is dominated by ionizing photons, while fluid on the other side of the interface is neutral and cold, and the radiation there is shifted to non-ionizing frequencies. If radiation pressure forces dominate gas pressure ones in the ionized gas, this gas is swept into a thin atmosphere on the surface of the front. The downconversion of the ionizing radiation to non-ionizing frequencies occurs mostly within this thin transition region \citep{KrumholzMatzner09, Draine2010}, and thus we can treat the situation as an interface problem.\footnote{Strictly speaking such an interface has a flow across it, since the amount of ionized mass increases with time in such a configuration. However, for a strong $D$ type ionization front the flux of mass and momentum across the ionized-neutral interface is very small compared to the flux reaching the front, and so we may safely neglect it.}

In this limit, it is most convenient to lump the gravitational and radiation forces in equation (\ref{momentum-gas}) together as:
\begin{equation}
\rho\mathbf{g}+\mathbf{G}_0=\rho\left(\mathbf{g}+\frac{\kappa_F}{c}\mathbf{F}\right),
\end{equation}
where we have ignored the difference between the comoving and the reference frame, as appropriate in this regime in the nonrelativistic limit. If, as we shall henceforth assume, the specific opacity $\kappa_F$ does not depend upon density, these two forces are exactly equivalent to an effective gravity field $\mathbf{g}_{\mathrm{eff}}\equiv \mathbf{g}+(\kappa_F/c)\mathbf{F}=-g_{\rm eff}\mathbf{e}_z$ constant in each medium, but which may differ, as mentioned above, between the two media.

  The equilibrium density profile on both sides of the interface is $\rho \propto \exp{(-g_{\rm eff} z/a^2)}$, where $a$ is the sound speed, so the scale height is $a^2/g_{\rm eff}$.

\subsection{Stability analysis}

We now move on to the derivation of the dispersion relation and the instability criterion. The underlying hydrodynamic equations are the same as in the classical Rayleigh-Taylor case, except that we replace $g$ by $g_{\rm eff}$, and we relax the assumption of incompressibility. The perturbed equations analogous to (\ref{incompressibility-perturbed-classical}) - (\ref{Eulerx-perturbed-classical}) in this case are
\begin{eqnarray}
\label{continuity-perturbed}
-i\omega\, \delta\rho+\delta v_z \frac{\partial}{\partial z}\rho+\rho\left(ik\delta v_x+\frac{\partial}{\partial z}\delta v_z\right) & = & 0 \\
-i\omega\rho \, \delta v_z + \frac{\partial}{\partial z}\delta P_g - g_{\rm eff} \, \delta \rho & = & 0\\
-i\omega \rho\, \delta v_x + i k \, \delta P_g & = & 0
\end{eqnarray}
Eliminating $\delta v_x$ as in \S 3, using the isothermal equation of state $P_g = \rho a^2$, and using the fact that $\partial \rho/\partial z = -(g_{\rm eff}/a^2)\rho$ and  for the background state, we obtain
\begin{equation}
\label{optically-thin-system}
\frac{d}{dz}
\left[\begin{array}{r}
\hat{\xi}_z\\\frac{\delta\hat{\rho}}{\rho}
\end{array}\right]
=
A
\left[\begin{array}{r}
\hat{\xi}_z\\\frac{\delta\hat{\rho}}{\rho}
\end{array}\right],
\end{equation}
with:
\begin{equation}
A=
\left[\begin{array}{cc}
\frac{g_{\mathrm{eff}}}{a^2} & \left(\frac{ka}{\omega}\right)^2-1\\
\left(\frac{\omega}{a}\right)^2&0 \\
\end{array}\right]
..
\label{A isothermal}
\end{equation}

The matrix $A$ here is independent of $z$ in each medium (as in \S 3). The general solution will thus adopt the form of equation (\ref{general-solution}). Hence, we need to discuss the eigenvalues of $A$. They satisfy the characteristic equation
\begin{equation} 
\lambda^2-\frac{g_{\mathrm{eff}}}{a^2}\lambda+\left(\frac{\omega}{a}\right)^2-k^2=0
\label{characteristic-isothermal}
\end{equation}
Let us focus our attention to medium 1. For the velocity \textit{and} density perturbation to vanish for $z\rightarrow +\infty$, we require that, for each eigenvector of $A$ with corresponding eigenvalue $\lambda_1$ along which the solution has a nonzero projection,
\begin{equation}
\mathrm{Re}(\lambda_1) < \mathrm{min}\left(0, \frac{g_1}{a_1^2}\right),
\end{equation}
where $g_1$ and $a_1$ are the value of $g_{\rm eff}$ and $a$ in medium 1. However, from the characteristic equation, we know that the average of the real parts of the two eigenvalues is $g_1/2a_1^2$, and therefore one of the eigenvalues has a real part that violates the above inequality, regardless of the sign of $g_1$. Therefore, the solution cannot have a nonzero projection along this eigenvector, and \textit{must} be an eigenvector of $A$. The same argument can be repeated in region 2 (one can e.g.\ change the orientation of the $z$ axis to be in the exact same configuration) and thus, $\delta\hat{\psi}$ (which is \textit{not} continuous at the interface) is an eigenvector of $A$ in each region, of eigenvalue $\lambda_1$ and $\lambda_2$.

To derive the dispersion relation, we now introduce the boundary condition at the interface. Equation (\ref{gas-pressure-continuity}) continues to hold if we replace $g$ with $g_{\rm eff}$, so
\begin{equation}
\left[\Delta P_g\right]^1_2=\left[\delta P_g - \rho g_{\rm eff} \xi_z\right]^1_2=0.
\end{equation}
Applying this to the characteristic equation (\ref{characteristic-isothermal}), we have
\begin{equation}
\frac{1}{a_1^2}\left(\frac{\omega^2}{\lambda_1}-g_1\right)=\frac{1}{a_2^2}\left(\frac{\omega^2}{\lambda_2}-g_2\right),
\label{pressure-continuity-isothermal}
\end{equation}
If we divide equation (\ref{characteristic-isothermal}) by $\lambda^2$ and equate the resulting left-hand-sides for each medium, use of equation (\ref{pressure-continuity-isothermal}) yields (since $\lambda_1\neq\lambda_2$)
\begin{equation}
\lambda_1\lambda_2=-k^2
\label{product-eigenvalues},
\end{equation}
which implies the two eigenvalues have real parts of opposite signs. While the constraint that the velocity perturbation vanishes at $+\infty$ and $-\infty$ hereby reduces to $\mathrm{Re}(\lambda_1)<0$, this does not guarantee that the density perturbation will do so in a medium where $\mathbf{g_{\mathrm{eff}}}$ points away from the interface. In this case, one needs to further satisfy\footnote{We start from the inequality $\mathrm{Re}(\lambda_1)<g_1/a_1^2$ (if we take the medium in question to be medium 1), with $\lambda_1$ obtained from solving the quadratic equation (\ref{characteristic-isothermal}). We use the following useful relationships, holding for all complex values of $z$:
$\mathrm{Re}(\sqrt{z})=\sqrt{\left(|z|+\mathrm{Re}(z)\right)/2}$ and $\mathrm{Im}(\sqrt{z})=\mathrm{sgn}(\mathrm{Im}(z))\sqrt{\left(|z|-\mathrm{Re}(z)\right)/2}$; the latter equation also defines our choice of branch cut in the complex plane.}
\begin{equation}
\left(\mathrm{Im}(\omega^2)\right)^2+g_{\mathrm{eff}}^2\left(k^2-\frac{\mathrm{Re}(\omega^2)}{a^2}\right)>0.
\label{if geff away}
\end{equation}
Solving equation (\ref{product-eigenvalues}) for $\lambda_2$ and injecting into equation (\ref{pressure-continuity-isothermal}), one obtains the following quadratic equation:
\begin{equation}
\left(\frac{a_1\lambda_1}{k}\right)^2-\frac{h}{\omega^2}\lambda_1+a_2^2=0,
\label{intermediary quadratic}
\end{equation}
with $h\equiv g_1a_2^2-g_2a_1^2$, from which one deduces that $\mathrm{sgn}(\mathrm{Re}(\lambda_1))=\mathrm{sgn}\left((\rho_1g_1-\rho_2g_2)\mathrm{Re}(\omega^2)\right)$ (with $\rho_{1,2}=\rho(0^\pm)$). Therefore, if $\rho_1g_1>\rho_2g_2$, $\omega$ cannot be real and the configuration is unstable. Indeed, even if $\mathrm{Im}(\omega)<0$, we simply need to take the complex conjugate of equations (\ref{optically-thin-system})-(\ref{A isothermal}): we then see that $\delta\hat{\psi}^\ast$ corresponds to a perturbation (still satisfying the boundary conditions) with the same wavenumber $k$ but with complex frequency $\omega^\ast$, and which is therefore a growing eigenmode. In a more coordinate-free manner (since upper and lower are not well-defined if the direction of the effective gravity switches sign across the interface), the instability arises if the effective weight per unit volume toward the interface overcomes that away from it. A corollary is that, should $\mathbf{g_{\mathrm{eff}}}$ point toward the interface in both regions, the equilibrium is unequivocally unstable.

\begin{figure}
\plotone{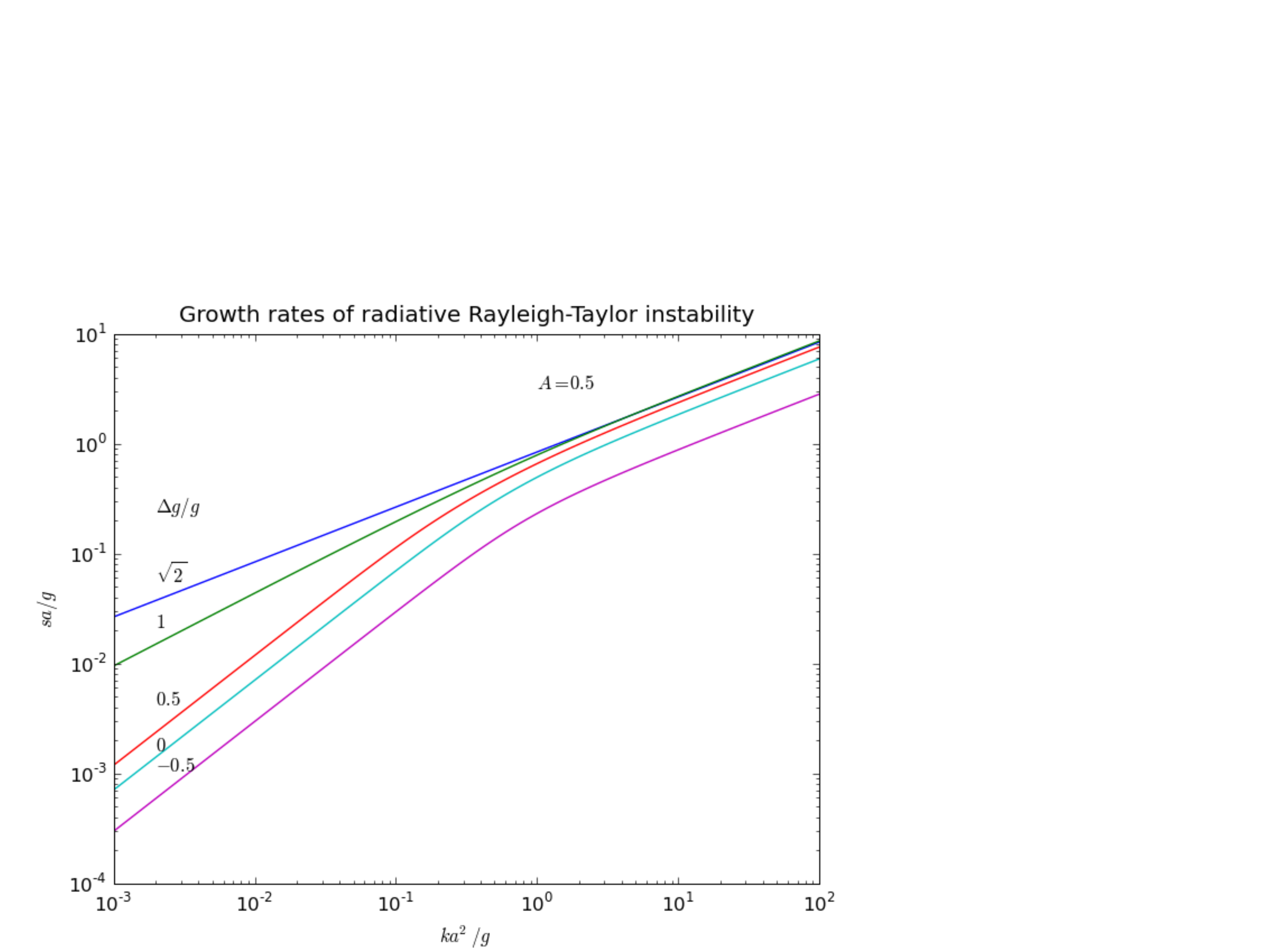}
\caption{Growth rate $s\equiv {\rm Im}(\omega)$ of the isothermal, optically-thin radiative Rayleigh-Taylor instability, calculated numerically from the dispersion relation (\ref{dispersion-isothermal}). The value shown corresponds to the fastest growing mode, i.e.\ to the largest value of $s$. Growth rates and wavenumbers are nondimensionalized through combinations of $a\equiv(a_1^2+a_2^2)^{1/2}$ and $g\equiv(g_1^2+g_2^2)^{1/2}$, and as such depend on two dimensionless parameters $\Delta g/g=(g_1-g_2)/g$ and the Atwood number $A=(\rho_1-\rho_2)/(\rho_1+\rho_2)$ (plus sign information, $\mathrm{sgn}(g_1+g_2)$) which are bounded by $2^{1/2}$ and 1 in absolute value, respectively. Here, we have fixed $A=0.5$ (and $g_1+g_2>0$) and varied $\Delta g/g$ with values (from top to bottom) $2^{1/2}$, $1$, $0.5$, $0$ and $-0.5$}
\label{RT-isothermal-growth}
\end{figure}

In order to completely prove the sufficiency of the criterion, we need to show that $\mathrm{Re}(\omega^2)<0$ (which ensures the inequality (\ref{if geff away})) is actually allowed by the dispersion relation. To do so, we first solve for $\lambda_1$ from equation (\ref{intermediary quadratic}) after eliminating the quadratic term with equation (\ref{characteristic-isothermal}). We obtain
\begin{equation}
\lambda_1=\omega^2\frac{\omega^2-(ka)^2}{g_1\omega^2-hk^2},
\end{equation}
with $a\equiv\sqrt{a_1^2+a_2^2}$. Since the same formula holds for $\lambda_2$ if the subscripts ``1'' and ``2'' exchange roles (and thus $h$ switches sign), the equation (\ref{product-eigenvalues}) yields the desired dispersion relation
\begin{eqnarray}
0 & = & \omega^8-2(ka)^2\omega^6+\left[(ka)^4+g_1g_2k^2\right]\omega^4
\nonumber\\
& & {}+k^4(g_1-g_2)h\omega^2-k^6h^2,
\label{dispersion-isothermal}
\end{eqnarray}
which is fourth order in $\omega^2$. As this polynomial always has one negative real root (in terms of $\omega^2$), provided $kh\neq 0$, the sufficiency of the instability criterion $\rho_1g_1>\rho_2g_2$ (or, equivalently, $h>0$) is proven. The polynomial has two real roots, only one of which is physically allowed (the other being opposite to leading order), asymptotically given by
\begin{eqnarray}
\omega^2 & =&-\frac{hk}{a^2}+\frac{h(g_1+g_2)(a_2^2-a_1^2)}{2a^6}+O\left(\frac{1}{k}\right)
\nonumber\\
&=&- k\frac{\rho_1g_1-\rho_2g_2}{\rho_1+\rho_2}
\nonumber \\
& & {}
+\frac{\left(g_1+g_2\right)\left(\rho_2-\rho_1\right)\left(\rho_1g_1-\rho_2g_2\right)}{2a^2\left(\rho_1+\rho_2\right)^2}
\nonumber\\
& & {} +O\left(\frac{1}{k}\right).
\label{isothermal-approximate}
\end{eqnarray}
In the long-wavelength limit, if the instability criterion is satisfied, $\omega^2$ is given by $(hk)^2\mathrm{min}\left(1/g_1,-1/g_2\right)$ if $g_1g_2>0$ and $-\sqrt{-g_1g_2}k$ if $g_1g_2<0$. 
The instability is a ``pure'' instability (in the sense that $\omega$ is purely imaginary). Figure \ref{RT-isothermal-growth} shows a calculation of the growth rate as a function of various parameters. Note that equation (\ref{isothermal-approximate}) agrees with earlier results for compressible Rayleigh-Taylor instability without radiation (equation 23 of \citealt{Shivamoggi08}) if we take $g_1 = g_2$.

\subsection{Sample application: radiation pressure-driven HII regions}

We conclude this section with a sample application for the case of the ionization front around an H~\textsc{ii} region where radiation pressure significantly affects the dynamics (e.g.\ the 30 Doradus region, \citealt{Lopez10a}). Consider such a region powered by a star cluster of luminosity $L_*$ expanding into a uniform ambient medium of number density $n$, sweeping up ambient gas as it expands. We will neglect the gravitational pull of the star cluster, which is significant only early in the evolution. During the radiation-dominated phase of the expansion, which applies when the H~\textsc{ii} radius $r\ll r_0$, the radius of the H~\textsc{ii} after a time $t$ is \citep{KrumholzMatzner09}
\begin{equation}
r \approx r_0 (t/t_0)^{1/2},
\end{equation}
where $r_0 = 11 L_7$ pc, $t_0 = 50 L_7^{3/2} n_6^{1/2}$ Myr, $L_7 = L_*/10^7$ $L_\odot$, $n_6 = n/10^6$ H nuclei cm$^{-3}$, and we for all other quantities have adopted the fiducial parameters of \citeauthor{KrumholzMatzner09}\footnote{except that we take $\psi = 3.3$, and we correct a factor of 2.2 error in equation 4 of Krumholz \& Matzner -- see \citet{Fall10a} for details} for the embedded case. Our choice of luminosity and density are motived by the example of the 30 Doradus H~\textsc{ii} region, which is driven by a central cluster of luminosity $1.7\times 10^7$ $L_\odot$.

The acceleration of the shell, and thus the effective gravitational force toward the front in the frame comoving with the front, is
\begin{eqnarray}
g_1 & = &
-\frac{r_0}{4t_0^2} \left(\frac{t}{t_0}\right)^{-3/2} 
\nonumber \\
& = & -1.1 \times 10^{-12} \, L_7^{-2} n_6^{-1} \left(\frac{t}{t_0}\right)^{-3/2}\mbox{ cm s}^{-2},
\end{eqnarray}
where positive sign corresponds to $\mathbf{g}_{\rm eff}$ pointing toward the cluster. Note that, for simplicity we have assumed that the material outside the transition region near the front is optically thin to the (non-ionizing) radiation that emerges from inside it. Within the shell, the opacity is dominated by dust (except near the transition region, where the neutral fraction is high enough for neutrals to contribute significantly). Radiation exerts a force toward the front, inducing an outward force per unit mass in the frame of the front given by 
\begin{eqnarray}
g_2 & = & g_1-\frac{\kappa_F L}{4\pi r^2 c} \\
& = & -8.8\times 10^{-8} \kappa_3 L_7^{-1} \left(\frac{t}{t_0}\right)^{-1/2} \mbox{cm s}^{-2},
\end{eqnarray}
where $\kappa_3 = \kappa_F / 10^3$ cm$^2$ g$^{-1}$, and in the numerical evaluation we have dropped $g_1$ since it is small compared to $g_2$. The normalization of $\kappa_F$ is chosen because $\sim 10^3$ cm$^2$ g$^{-1}$ is a typical dust opacity for radiation at the color temperature of an O star \citep{draine03}.

If we adopt sound speeds of $a_1 = 0.19$ km s$^{-1}$ and $a_2 = 9.2$ km s$^{-1}$ in the neutral and ionized gas (appropriate for molecular gas at 10 K and ionized gas at $7000$ K, respectively), $\rho_1g_1-\rho_2g_2=(\rho_1-\rho_2)g_1+\rho_2\kappa_FF/c\approx \rho_2\kappa_FF/c >0$, and we find that this configuration is unstable. The inertial force term $\rho_1 g_1$ points away from the interface and is therefore stabilizing, and instability occurs only because it is overcome by the larger $\rho_2 g_2$ term that is dominated by radiation force. In the short wavelength limit, which applies for all perturbations smaller than $r_0$, the growth rate is given by (using equation \ref{isothermal-approximate}) 
\begin{equation}
{\rm Im}(\omega) \approx \frac{\sqrt{hk}}{a} \approx \frac{a_1}{a_2} \sqrt{g_2 k}.
\end{equation}
Plugging in our fiducial values, we have
\begin{equation}
{\rm Im}(\omega) \approx 0.08 \left(\frac{\kappa_3}{L_7}\right)^{1/2} \left(\frac{t_0}{t}\right)^{1/4} \left(\frac{r_0}{\lambda}\right)^{1/2} \mbox{ Myr}^{-1},
\end{equation}
and we learn that modes with $\lambda/ r_0 \la 0.01$ ($\lambda \la 0.1$ pc for our fiducial parameters) will be able to grown significantly in the few Myr lifetime of the stars driving the H~\textsc{ii} region. This may explain the small-scale filamentary structures seen around the edges of 30 Doradus and similar radiatively-driven H~\textsc{ii} regions.

\section{The ``adiabatic'' Rayleigh-Taylor instability}

\subsection{Formulation of the equations}

We now consider the stability of an interface where both sides are optically thick (although we discuss how to relax this requirement for the lower medium below) and in radiative equilibrium. In this regime, the radiation field is a Planckian locked at the gas temperature and the comoving frame pressure tensor may be taken to be isotropic (scalar) and given by $\mathcal{P}_{r0}=(E_{r0}/3)\mathcal{I}_3$, with $E_{r0}=a_RT^4$. Equation (\ref{momentum-radiation-comoving}) may then be approximated by
\begin{equation}
\mathbf{G_0}\approx-\nabla P_{r0},
\end{equation}
which can be lumped with the gas pressure force. In making this assumption, we require that the photon mean free path $1/(\kappa_F\rho)$ be smaller than the wavelengths of the perturbation and the characteristic lengthscale of variation of the background equilibrium. The latter may be defined as $L\equiv\mathrm{min}\left(P_{\mathrm{tot}}/\rho g,P_{r0}c/\kappa_FF\right)=\mathrm{min}\left(1+x,x/E\right)a^2/g$, where for convenience we define
\begin{eqnarray}
E & \equiv& \frac{\kappa_F F_0}{gc} \\
x & \equiv & \frac{P_{r0}}{P_g}.
\end{eqnarray}
Physically, $E$ measures the Eddington ratio of the background state (i.e.\ the ratio of radiation force to gravitational force), while $x$ measures the relative importance of radiation and gas pressure.

Considerable simplification of the problem is achieved if we are allowed to drop $\nabla\cdot\mathbf{\delta F_0}$ in the energy equation. (Appendix \ref{optically-thick-appendix} shows the system of equations without this approximation.) Indeed, the energy equation (\ref{total-internal-energy}) can be rewritten as
\begin{equation}
\rho T\frac{Ds}{Dt}=-\nabla\cdot\mathbf{F_0},
\label{2nd principle}
\end{equation}
with $s$ the specific entropy of the gas plus radiation fluid,
\begin{equation}
 s=\frac{k_B}{m(\gamma-1)}\mathrm{ln}P_g\rho^{-\gamma}+\frac{4P_{r0}}{\rho T}
\end{equation}
 Perturbation of equation (\ref{2nd principle}) yields
\begin{equation}
\frac{D\Delta s}{Dt}=-\frac{\nabla\cdot\delta\mathbf{F_0}}{\rho T}
\end{equation}
If we can actually disregard the right-hand-side, we obtain
\begin{equation}
 \Delta s=0.
\label{adiabatic}
\end{equation}
Such an approximation, henceforth referred to as the ``adiabatic approximation'', holds in the limit of high optical thickness. More precisely, when discussing the validity of the upcoming calculation, we will require
\begin{equation}
\delta F_0\ll E_{\mathrm{tot}}\delta v.
\label{validity-adiabatic}
\end{equation}
A possible astrophysical realization of this configuration is the wall of a rarefied bubble blown by a massive star in formation \citep[e.g.][]{Krumholzetal2009}. The system is close to the Eddington limit, so the gravity and luminosity of the central star (at $z=-\infty$) nearly balance. Medium 1 would correspond to dust-laden gas infalling from the protostellar core while medium 2 would refer to the rarefied bubble. Flow of matter across the interface is very slow and can therefore be neglected.

\subsection{Stability analysis}

While the perturbed mass conservation equation is the same as in \S 4 (equation (\ref{continuity-perturbed})), the two components of the perturbed Euler equation read
\begin{eqnarray}
-i\omega\rho\delta v_z & = & -\frac{\partial}{\partial z}\delta P_{\mathrm{tot}}-g\delta\rho
\label{Eulerz-perturbed-adiabatic}
\\
-i\omega\rho\delta v_x & = &-ik\delta P_{\mathrm{tot}}
\label{Eulerx-perturbed-adiabatic}
\end{eqnarray}
The total pressure continuity at the interface is
\begin{equation}
\left[\Delta P_{\mathrm{tot}}\right]^1_2=\left[\delta P_g + \delta P_{r0} - \rho g \xi_z\right]^1_2=0.
\label{pressure-continuity-adiabatic}
\end{equation}
We eliminate $\delta v_x$ through equation (\ref{Eulerx-perturbed-adiabatic}) as previously and $\delta\rho$ through the following formula (from the expression of $s$):
\begin{equation}
\frac{\delta\rho}{\rho}=\left(C\frac{\delta P_{\mathrm{tot}}}{P_g}-\frac{1+4x}{D}\frac{m}{k_B}\delta s\right),
\end{equation}
combined with equation (\ref{adiabatic}). Here,
\begin{eqnarray}
\label{eq:cdef}
C & = & \frac{1}{D}\left(12x+\frac{1}{\gamma-1}\right) \\
\label{eq:ddef}
D & =& 16x^2+20x+\frac{\gamma}{\gamma-1}.
\end{eqnarray}

The ODE system in $z$ then reduces to a $2\times 2$ matrix $A$, defined by
\begin{equation}
\frac{d}{dz}\left[\begin{array}{c}
\hat{\xi}_z\\\delta \hat{P}_{\mathrm{tot}}
\end{array}\right]
=A
\left[\begin{array}{c}
\hat{\xi}_z\\\delta\hat{P}_{\mathrm{tot}}
\end{array}\right],
\end{equation}
with
\begin{equation}
A
=
\left[\begin{array}{cc}
\frac{g}{a^2}C&\frac{1}{\rho}\left(\left(\frac{k}{\omega}\right)^2
-\frac{C}{a^2}\right)\\
\rho\left(\omega^2+B\left(\frac{g}{a}\right)^2\right)
&-\frac{g}{a^2}C \\
\end{array}\right],
\end{equation}
where
\begin{eqnarray}
 B& =& \frac{1}{D}\bigg(16(E-1)x^2+(24E-8)x
 \nonumber\\
 &&
  {}+E(5+\frac{\gamma}{\gamma-1})-1+\frac{\gamma E}{4(\gamma-1)x}\bigg)
.
\label{eq:bdef}
\end{eqnarray} 

Let us restrict attention to eigenmodes actually localized at the interface, i.e., which vanish on a vertical lengthscale small compared to $L$. This we will refer to as the ``evanescence condition''. Then, the matrix $A$ above may be treated as a constant in each region (we provisionally consider the adiabatic approximation to hold in the lower region too). Since it is traceless (and thus the two eigenvalues are equal and opposite), $\hat{\psi}$ must be an eigenvector of $A$ (as it was in the previous section) in each region, of eigenvalue $\lambda_1$ and $\lambda_2$ in media 1 and 2, where the eigenvalue have negative and positive real parts, respectively.

Combining the first row of the eigenvalue equation for $z=0^\pm$ and the continuity of the pressure, one finds that
\begin{equation}
\left[\rho\frac{\lambda\left(\omega/k\right)^2-g}{1-C\left(\omega/ka\right)^2}\right]^1_2=0.
\label{adiabatic-dispersion1}
\end{equation}
More explicitly, the dispersion relation implied by the above reads
\begin{eqnarray}
\frac{\rho_1}{1-C_1\left(\omega/ka_1\right)^2}\bigg(-g-\left(\frac{\omega}{k}\right)^2\nonumber\\\sqrt{-\frac{C_1}{a_1^2}\omega^2-\frac{C_1(B_1-C_1)g^2}{a_1^4}+k^2+B_1\left(\frac{kg}{a_1\omega}\right)^2}\bigg)\nonumber\\
=\frac{\rho_2}{1-C_2\left(\omega/ka_2\right)^2}\bigg(-g+\left(\frac{\omega}{k}\right)^2\nonumber\\\sqrt{-\frac{C_2}{a_2^2}\omega^2-\frac{C_2(B_2-C_2)g^2}{a_2^4}+k^2+B_2\left(\frac{kg}{a_2\omega}\right)^2}\bigg).
\label{adiabatic-dispersion2}
\end{eqnarray}
This equation may (in principle) be manipulated to achieve polynomial form, but with a degree of 20, and a loss of sign information. 

We now further specialize to the case where $\rho_2\ll\rho_1$. Equation (\ref{adiabatic-dispersion1}) then reduces to
\begin{equation}
\lambda_1=g\left(\frac{k}{\omega}\right)^2,
\label{lambda-adiabatic}
\end{equation}
This equation we would have obtained more directly had we set $\Delta_1 P_{\mathrm{tot}}=\Delta_2 P_{\mathrm{tot}}=0$ and thus its validity is not endangered if the adiabatic approximation is violated in medium 2, or if the latter is optically thin, provided we consider this rarefied medium to have an imposed radiation field. We observe that $\Delta P_{\mathrm{tot}}=0$ combined with $\Delta s=0$ entails $\Delta T=0$ and $\Delta\rho=0$ \textit{at the interface}. 

From equation (\ref{lambda-adiabatic}), one deduces that we must have $\mathrm{Re}(\omega^2)<0$ if $g>0$, in order for the perturbation not to blow up for $z\rightarrow+\infty$, and thus, we have instability. Dropping henceforth the '1' subscripts, the evanescence condition then translates to
\begin{equation}
\frac{\omega}{k}<a\sqrt{\mathrm{min}\left(1+x,\frac{x}{E}\right)}.
\label{evanescence-explicit}
\end{equation}
The dispersion relation follows from setting the right-hand-side of equation (\ref{adiabatic-dispersion2}) to zero:
\begin{equation}
-\frac{C}{a^2}\omega^6+\left(\frac{C(C-B)g^2}{a^4}+k^2\right)\omega^4+B\left(\frac{kg\omega}{a}\right)^2-g^2k^4=0,
\label{dispersion-adiabatic-order6}
\end{equation}
which is third order in $\omega^2$ and always has a negative root (in terms of $\omega^2$), whence the instability qualifies as a pure instability.

In the long-wavelength limit, the complex frequency converges toward a finite value given by
\begin{eqnarray}
\omega^2 & = & \left(\frac{g}{a}\right)^2\left(C-B\right)
\nonumber\\
& = &\left(\frac{g}{a}\right)^2\left\{1-E\left[1+\frac{4x+5+\frac{\gamma}{4(\gamma-1)x}}{16x^2+20x+\frac{\gamma}{\gamma-1}}\right]\right\},
\end{eqnarray}
which amounts to a Brunt-Va\"{i}sala frequency modified by compressibility, corresponding to a positive growth at (and even somewhat below) the Eddington limit. Indeed, the background entropy gradient
\begin{equation}
\frac{ds}{dz}=\frac{g}{T}\left\{1+4x-E\left[5+4x+\frac{\gamma}{4(\gamma-1)x}\right]\right\}
\label{dzs}
\end{equation}
is already negative at the Eddington limit, since a large radiative flux corresponds to a large temperature gradient in the optically thick limit\footnote{Since $F_0<E_{r0}c$, we must have $E/x<\kappa_FP_g/g$, and by a large margin at that if the diffusion approximation is to hold.}. However, as we shall see in the next section, this long-wavelength limit often violates the evanescence condition (except for a nonzero range if $B\,\mathrm{min}\left(1+x,x/E\right)\gg 1$), in which case this convective instability, although physically sensible because of equation (\ref{dzs}), is not captured quantitatively by the present calculation.

Since the evanescence condition allows one to neglect the highest order term and the first term inside the parentheses of the second one in equation (\ref{dispersion-adiabatic-order6}), we can provide the following more explicit (approximate) formula: 
\begin{equation}
\omega^2=g\left(-\frac{Bg}{2a^2}-\sqrt{k^2+\left(\frac{Bg}{2a^2}\right)^2}\right)
\end{equation}
For large wavenumbers, this coincides with the standard Rayleigh-Taylor result, and indeed, physically, the instability in question is really Rayleigh-Taylor in essence: this results from the fact that, in the adiabatic approximation, the system is completely mathematically equivalent to a single fluid with a modified equation of state (total pressure law). We emphasize that this ``adiabatic Rayleigh-Taylor instability'', as we will call it for ease of reference, is \textit{not} an interface counterpart of the local RHD instability studied by \citet{BlaesSocrates2003}, where acoustic disturbances are overstable, whereas perturbations are here incompressible at the interface. Moreover, the instability criterion of \citet{BlaesSocrates2003} involves the opacity law, while opacity is actually absent in the equations under the adiabatic approximation, except as a constraint for their validity. The domain of validity of the above calculation we shall now discuss.

\subsection{Validity of the adiabatic approximation}

\begin{figure}
\plotone{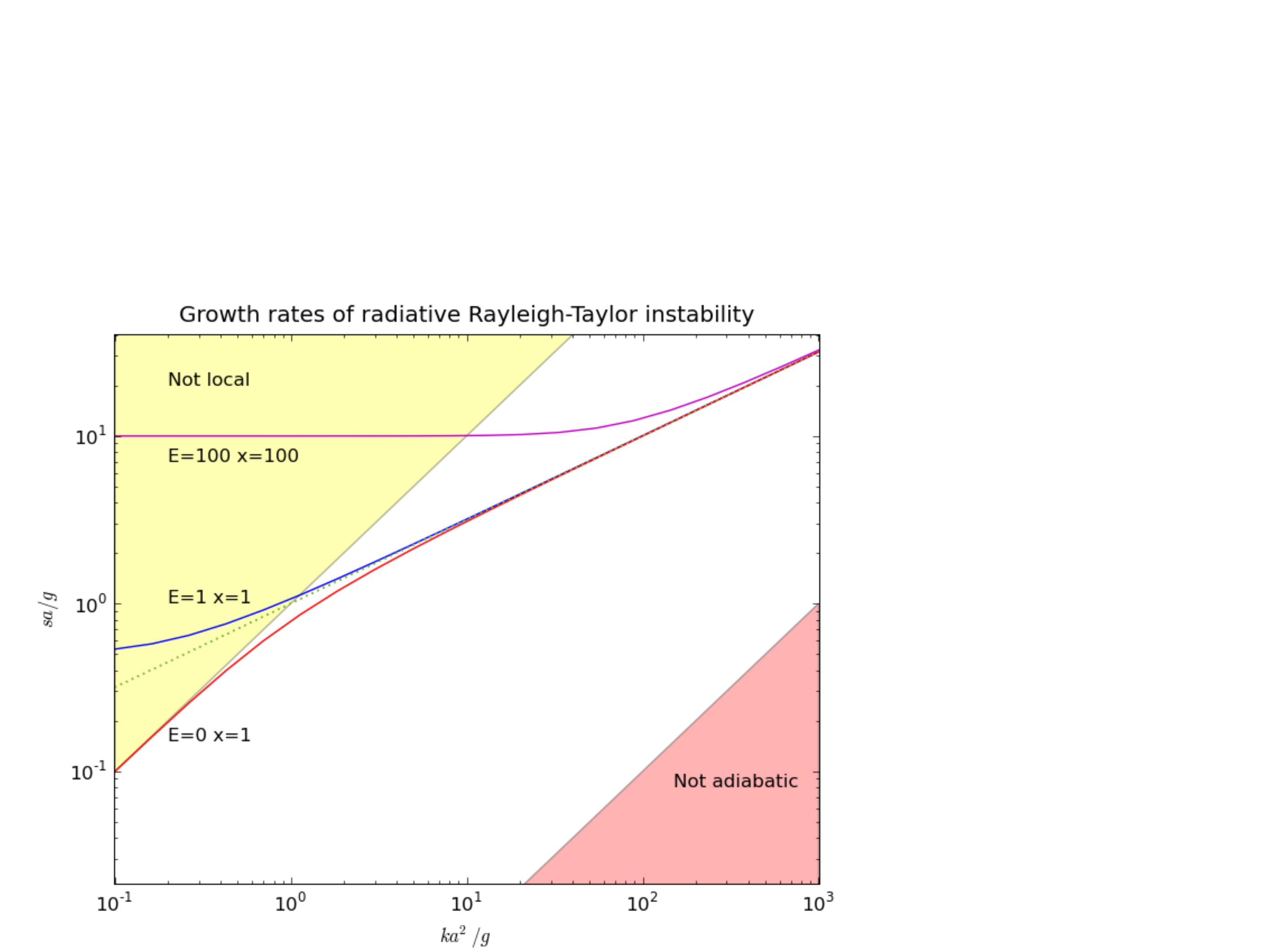}
\caption{Growth rate $s\equiv {\rm Im}(\omega)$ of adiabatic Rayleigh-Taylor instability, calculated numerically from the full dispersion relation (\ref{dispersion-adiabatic-order6}) (solid line) with the classic Rayleigh-Taylor result overplotted (dotted line). Shaded in yellow is the region of the k-s plane where the evanescence constraint (the decay of perturbations on lengthscales smaller than the background equilibrium scale height) is violated, and in red that of breakdown of the adiabatic approximation (conservation of specific entropy for gas plus radiation) breaks down, for a given value of $F_0/E_{\mathrm{tot}}a$ (which is here arbitrary, since the calculation does not depend on $F_0$). For completeness, we note that allowable growth rates and wavenumbers are also bounded by radiative equilibrium and optical thickness requirements, respectively. We have plotted the growth rates for three representative sets of parameters $E=0$, $x=1$ (for which the growth rate vanish in the long wavelength limit) $E=x=1$ and $E=x=100$ (for which the growth rate converges toward a finite value in the long wavelength limit). In the short wavelength limit, the growth rates are identical to the classical Rayleigh-Taylor value.}
\label{RT-adiabatic-growth}
\end{figure}  

Since $\delta P_{r0}=-\xi_z (dP_{r0}/dz)=E\rho g\xi_z $, the requirement (\ref{validity-adiabatic}), which amounts to $kc\, \delta P_{r0}/\kappa \rho \ll E_{\mathrm{tot}}\delta v_z$,
is equivalent to
\begin{equation}
\frac{\omega}{k}\gg \frac{F_0}{E_{\mathrm{tot}}}
\label{adiabatic-validity-explicit}
\end{equation}
Figure \ref{RT-adiabatic-growth} shows the growth rate calculated for example parameters, with regions forbidden by the evanescence and the adiabatic constraints shaded. As mentioned previously, we see that the long-wavelength limit violates the evanescence condition, while at large enough wavenumbers ($k > g\left(E_{\mathrm{tot}}/F_0\right)^2\mathrm{min}\left(1+x,x/E\right)$ if we take $\omega^2\approx -gk$), the adiabatic approximation breaks down. 

Constraints (\ref{evanescence-explicit}) and (\ref{adiabatic-validity-explicit}) are \textit{compatible} (i.e., allow a range of wavenumbers in which the above calculation is valid) if
\begin{equation} 
F_0\ll E_{\mathrm{tot}}a\sqrt{\mathrm{min}\left(1+x,\frac{x}{E}\right)}
\label{compatibility}
\end{equation}
Inasmuch as it is a diffusion term that becomes important (in the energy equation) as the adiabatic approximation breaks down, it may be suspected that larger wavenumber disturbances are damped (yielding a maximum growth of order $\sqrt{\mathrm{min}\left(1+x,x/E\right)}gE_{\mathrm{tot}}/F_0=\sqrt{\mathrm{min}\left(1+x,x/E\right)}\kappa E_{\mathrm{tot}}/Ec$).

\subsection{Sample application: bubbles around massive protostars}

As a sample application, we consider radiation-driven bubbles around massive stars, such as those seen in the simulations of \citet{Krumholzetal2009} or \citet{Kuiper10}. In these simulations, the radiation emitted by an accreting massive star exerts a force stronger than gravity on the dusty gas around it that is trying to accrete. Above and below the midplane of the dense accretion disk, the radiation drives an expanding shell of material into the surrounding protostellar core. Inside the shell is an evacuated low-density region filled by radiation, while outside there is dense swept-up dusty gas. We are interested in exploring the stability of the interface between the low-density bubble and the high density shell, where the flow is near the Eddington limit. Given the relatively sharp edges of the radiation bubbles that form in the \citet{Krumholzetal2009} simulations, it does appear that they lend themselves to an interface stability analysis such as the one we present here. The continuous medium case treated by \citet{BlaesSocrates2003} indeed does not seem to be of relevance here, because the \citeauthor{Krumholzetal2009} simulations---which did not include magnetic fields---do not satisfy \textit{a priori} their local hydrodynamical instability criterion (equation 58, or 63 if one takes gas and radiation temperature to be equal) as their implemented specific Rosseland mean opacity is independent of gas density.

Consider a star of mass $M_*$ and luminosity $L_*$ that inflates a bubble of radiation to a stellocentric distance $R$. The material in the bubble wall has a density $\rho$ and a temperature $T$. The mean mass per particle is $m=2.33\: m_{\rm H}$ and the gas has $\gamma=7/5$, appropriate for warm molecular hydrogen. The bubble wall is defined by the condition $E\approx 1$, and we also have
\begin{equation}
x = \frac{a_R T^3 m}{3 \rho k_B} = 950\, T_s^3 \rho_{-16}^{-1},
\end{equation}
where $T_s = T/1100$ K (where 1100 K is the dust sublimation temperature in the simulations, and thus the temperature at the edge of the bubble) and $\rho_{-16} = \rho/10^{-16}$ g cm$^{-3}$, a typical density in the bubble wall in the simulations. First we can check that the compatibility condition (equation \ref{compatibility}) for application of the adiabatic approximation is satisfied. Doing so, we find
\begin{equation}
\frac{E_{\rm tot} a \sqrt{x}}{F_0} \approx 50 \frac{T_s^5 r_4^2}{L_5 \rho_{-16}^{1/2}},
\end{equation}
where $r_4 = r/10^4$ AU, $L_5 = L/10^5$ $L_\odot$, and we have chosen normalizations for our parameters based on typical bubble properties seen in the simulations. Since adiabaticity requires that this ratio be $\gg 1$, we see that the condition is satisfied, and instability is guaranteed since $\rho_2\ll \rho_1$.

\begin{figure}
\plotone{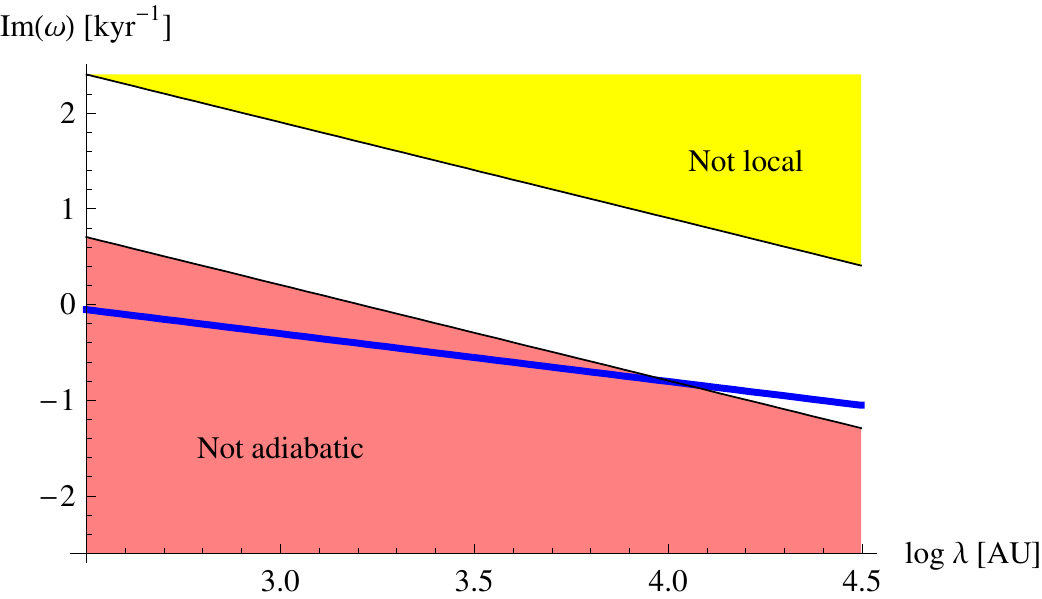}
\caption{
\label{highmass1}
Radiation Rayleigh-Taylor instability growth rate for the bubble around a massive star, as a function of the wavelength $\lambda=2\pi/k$ of the perturbation, computed for the parameters $T_s = 1$, $L_5 = 1$, $\rho_{-16} = 1$, $r_4 = 1$, and $M=100$ $M_\odot$. The blue line shows the computed growth rate, while the pink and yellow regions indicate where the conditions for the instability to be adiabatic (equation \ref{adiabatic-validity-explicit}) and evanescence (equation \ref{evanescence-explicit}) conditions are violated.
}
\end{figure}

\begin{figure}
\plotone{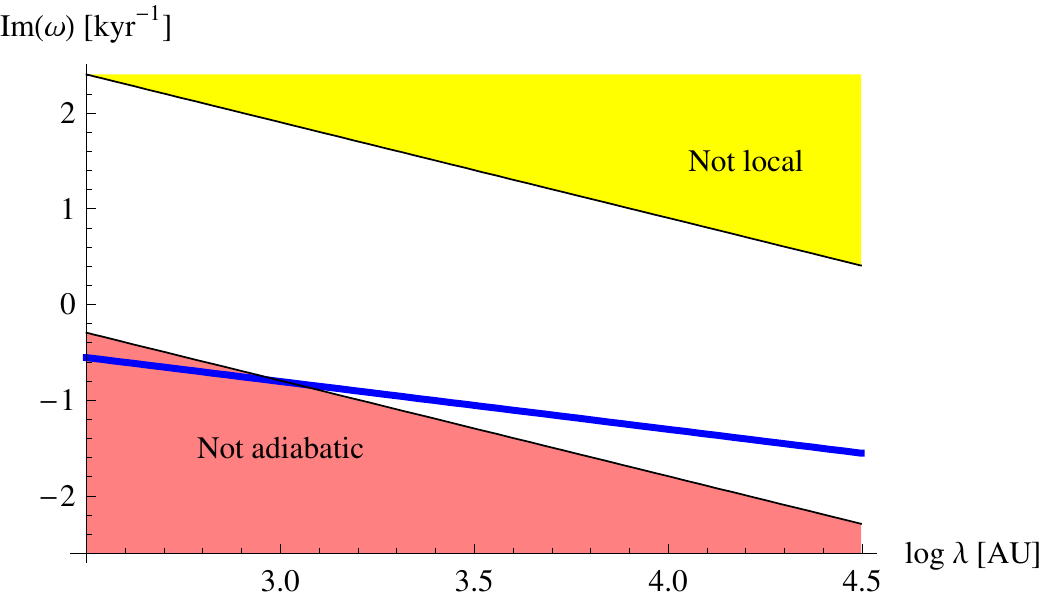}
\caption{
\label{highmass2}
Same as Figure \ref{highmass1}, but for the parameters $T_s = 1$, $L_5 = 0.1$, $\rho_{-16} = 1$, $r_4 = 1$, and $M=10$ $M_\odot$.
}
\end{figure}

To obtain the growth rate, we plug into equations (\ref{eq:cdef}), (\ref{eq:ddef}), and (\ref{eq:bdef}) for $B$, $C$, and $D$, and using 
 $g = G M_*/R^2$, and numerically solve the dispersion relation (\ref{dispersion-adiabatic-order6}) to find the fastest growing mode (i.e.\ the one with the largest negative value of $\omega^2$). We do so for two example sets of parameters in Figures \ref{highmass1} and \ref{highmass2}. In each case we see that the instability growth time is below 1 kyr, short compared to the $\sim 100$ kyr formation timescale for the star. At small wavelengths the instability is likely to be suppressed by diffusion, since the solution violates the adiabaticity constraint (\ref{adiabatic-validity-explicit}), but at large wavelengths (in some cases comparable to the physical size of the bubble) the constraint is satisfied and instability occurs. This instability explains the behavior observed in the \citeauthor{Krumholzetal2009}\ simulations, where large modes grow unstable and allow accretion onto massive stars that are formally super-Eddington. Although we have argued above that photon bubble instabilities were not relevant to interpret these simulations, this should not necessarily be extended to the ``real-world'' problem of massive star formation, e.g. because we have ignored magnetic fields (which have been detected observationally), unlike the local simulations of circumstellar enveloppes by \citet{turner07}. The relative importance of the photon bubble and the radiative Rayleigh-Taylor instabilities (as well as other processes, see e.g. \citealt{ZinneckerYorke2007}) for massive star formation has yet to be determined.

\section{Summary and conclusion}
We have studied the linear stability of a plane parallel superposition of two media separated by a thin interface, with both gravity and radiation force, and given results for two analytically tractable limiting cases. In these two cases, the role of radiation in these Rayleigh-Taylor-like instabilities is qualitatively different.

In the optically thin, isothermal limit, with a constant flux and a constant specific opacity in each medium, assumed to be chemically distinct, radiation acts like an effective gravitational field, which generally is different on either side of the interface. Linear instability occurs if the effective gravity per unit volume toward the interface overcomes that away from it, which in the case of a continuous effective gravity reduces to the ordinary Rayleigh-Taylor criterion on the Atwood number. This instability might contribute to the asymmetry of H~\textsc{ii} regions.

In the opposite limit, if the upper medium is optically thick and satisfies the approximation that the total specific entropy of the gas plus radiation fluid is conserved, assuming the lower medium to be rarefied, one finds that perturbations that vanish away from the interface more rapidly than the background equilibrium scale height are unstable. In the short-wavelength limit, the instability is indistinguishable from the classical Rayleigh-Taylor result, since the adiabatic approximation reduces the system to a single fluid, where the radiation force is part of the pressure force. Sufficiently close to the Eddington limit, the growth rate converges toward a finite value in the long-wavelength limit because of the negative entropy gradient. This instability could pertain to massive star formation by accretion beyond the Eddington limit \citep{Krumholzetal2009}.

Yet other regimes not studied in this work could be qualitatively different. For example, it is conceivable that the local radiative hydrodynamic overstabilities in optically thick media studied by \citet{BlaesSocrates2003} have an interface counterpart, where radiation slips into underdense regions near the interface. Those instabilities are not captured by the adiabatic approximation, and investigation of how the instability behaves beyond the adiabatic limit is left for future work. 

\acknowledgments
This project was initiated during the ISIMA 2010 summer program, funded by the NSF CAREER grant 0847477, the France-Berkeley fund, the Institute for Geophysics and Planetary Physics and the Center for Origin, Dynamics and Evolution of Planets. We thank them for their support. MRK acknowledges support from: an Alfred P.\ Sloan Fellowship; NSF grants AST-0807739 and CAREER-0955300; and NASA through Astrophysics Theory and Fundamental Physics grant NNX09AK31G and a Spitzer Space Telescope Theoretical Research Program grant.

\begin{appendix}

\section{The matrix $A$ in the optically thick regime without the adiabatic approximation}
\label{optically-thick-appendix}

The relevant perturbation equations are those mentioned in \S 4.2.1, plus the perturbed energy equation, diffusion approximation closure, and radiative flux continuity:
\begin{equation}
-i\omega\delta E_{\mathrm{tot}}+\delta v_z \frac{\partial}{\partial z} E_{\mathrm{tot}} + \nabla\cdot\mathbf{\delta F_0}+H_{\mathrm{tot}}\nabla\cdot\mathbf{\delta v}=0
\end{equation}
\begin{equation}
\mathbf{\delta F_0}=-\frac{c}{\chi}\nabla\delta P_{r0}-\mathbf{F_0}\delta \mathrm{ln}\chi
\end{equation}
\begin{equation}
\left[\delta F_{0,z}\right]^1_2=0,
\end{equation}
with $\chi\equiv\kappa_F\rho$. The matrix $A$ is $4\times 4$ and is defined by:
\begin{equation}
\frac{d}{dz}\left[\begin{array}{r}
\hat{\xi_z}\\\delta \hat{P}_{\mathrm{tot}}\\\delta\hat{P}_{r0}\\\delta\hat{F}_{0 z}
\end{array}\right]
=A
\left[\begin{array}{r}
\hat{\xi_z}\\\delta \hat{P}_{\mathrm{tot}}\\\delta\hat{P}_{r0}\\\delta\hat{F}_{0 z}
\end{array}\right],
\end{equation}
and given by:
\begin{equation}
A=
\left[\begin{array}{cccc}
-\frac{\partial}{\partial z}\mathrm{ln}\rho & \frac{1}{\rho}\left(\frac{k}{\omega}\right)^2-\frac{1}{P_g} & \frac{1}{P_g}+\frac{1}{4P_{r0}} & 0\\
\rho\omega^2 & -\frac{\rho g}{P_g} & \rho g \left(\frac{1}{P_g}+\frac{1}{4P_{r0}}\right) & 0\\
0 & -\frac{\chi F_{0 z}}{c}\frac{1+\Theta_{\rho}}{P_g} & \frac{\chi F_{0 z}}{c}\left((1+\Theta_{\rho})(\frac{1}{P_g}+\frac{1}{4P_{r0}})-\frac{\Theta_T}{4P_{r0}}\right) & -\frac{\chi}{c}\\
i\omega\left(\frac{\partial}{\partial z}E_{\mathrm{tot}}-H_{\mathrm{tot}}\frac{\partial}{\partial z}\mathrm{ln}\rho\right) & i\omega\left(\frac{1}{\gamma - 1}-\frac{H_{\mathrm{tot}}}{P_g}\right) & i\omega\left(3-\frac{1}{\gamma-1}+H_{\mathrm{tot}}(\frac{1}{P_g}+\frac{1}{4P_{r0}})\right)-\frac{k^2c}{\chi} & 0\\ 
\end{array}\right],
\end{equation}
with $\Theta_{\rho}\equiv\frac{\partial\mathrm{ln}\kappa_F}{\partial\mathrm{ln}\rho}_{|T}$ and $\Theta_T\equiv\frac{\partial\mathrm{ln}\kappa_F}{\partial\mathrm{ln}T}_{|\rho}$.
\end{appendix}

\bibliographystyle{apj}
\bibliography{bibliography}
\end{document}